\documentclass[prl,aps,twocolumn,epsfig]{revtex4}
\usepackage{graphicx}
\usepackage{dcolumn}
\usepackage{bm}

\begin{document}
\title {Resistance noise in electrically biased bilayer graphene}
\author{Atindra Nath Pal\footnote[1]{electronic mail: atin@physics.iisc.ernet.in}}
\author{Arindam Ghosh}
\address{Department of Physics, Indian Institute of Science, Bangalore 560 012, India}


\begin{abstract}
We demonstrate that the low-frequency resistance fluctuations, or noise, in
bilayer graphene is strongly connected to its band structure, and displays a
minimum when the gap between the conduction and valence band is zero. Using
double-gated bilayer graphene devices we have tuned the zero gap and charge
neutrality points independently, which offers a versatile mechanism to
investigate the low-energy band structure, charge localization and screening
properties of bilayer graphene.
\end{abstract}


\maketitle

The growing interest in bilayer graphene (BLG) is fueled by the ability to
control the energy gap between its valence and conduction bands through
external means \cite{mccann1,mccann2,castroneto}.  The Bernal stacking of
layers in BLG enforce strong interlayer coupling ($\gamma_1 \approx 0.35$~eV)
at the adjacent atomic sites, leaving two low-lying states that form a zero gap
semiconductor with quadratic dispersion. Setting finite potential difference
between the layers opens a gap $\Delta_g$ between the bands, which can be tuned
up to a maximum of $\approx \gamma_1$ using chemical doping \cite{castro,ohta},
or application of an external electric field \cite{oostinga}. Consequently,
biased BLG is not only attractive to nanoelectronics for device applications,
but also forms a new and versatile platform for studying wide range of
phenomena including magnetic instabilities \cite{stauber}, or weak localization
effects~\cite{savchenko,koshino} with massive chiral Dirac Fermions. However,
intrinsic disorder has been suggested to modify the band structure in BLG as in
conventional semiconductors by smearing the bands, and localizing the states in
band tails \cite{tailstate}. Most analysis assume disorder to be static, and
the influence of any kinetics or time-dependence of disorder on various
properties of BLG is still poorly understood.

Recently, the low frequency fluctuations, or the $1/f$ noise, in electrical
resistance of bilayer graphene has been shown to be sensitive to the BLG band
structure, in a way that is different from conventional time-averaged transport
\cite{Avouris}. The accepted mechanism of noise in graphene, as in carbon
nanotubes \cite{Avouris2}, is connected to potential fluctuations from the trap
states in the underlying silicon-oxide layers. In the case of monolayer
graphene, the noise magnitude decreases with increasing carrier density ($n$)
as the trap potentials are screened effectively by the mobile charges.
Conversely, noise in BLG {\it increases} with increasing $n$, thereby forming a
minimum around $n = 0$, which has been explained by the diminished ability of
BLG to screen the external potential fluctuations in presence of finite
$\Delta_g$. Indeed, theoretical models based on continuum self-consistent
Hartree potential approach~\cite{mccann2} or density-functional theory
calculations \cite{sahu} unanimously agree on maximal screening of the external
potential as $\Delta_g \rightarrow 0$. Thus, low-frequency noise in BLG
provides two crucial informations: (1) Inter-layer charge distribution in the
BLG since noise depends on the potential energy difference between the layers,
and (2) the nature of single-particle density-of-states (DOS) as well as the
chemical potential due to the sensitivity of noise to the underlying screening
mechanisms. However, in noise experiments on BLG so far~\cite{Avouris},
$\Delta_g$ has been tuned only by varying $n$ with a single (back) gate, where
a partial screening of the gate potential leads to excess charge in the upper
layer, and hence an electric field between the graphene layers \cite{mccann2}.
Here, we have measured the low-frequency resistance noise in spatially extended
double-gated BLG devices. The main objective is to achieve an independent
tunability of $\Delta_g$ with both $n$ and ${\cal E}$, where $\cal E$ is the
transverse electric field across the electrodes, to separate the influence of
band structure and carrier density on screening. Our experiments indicate that
multiple processes involving the charge traps are active in producing the
resistance noise which is intimately connected to the BLG band structure, being
minimum at $\Delta_g = 0$ even if it corresponds to a nonzero $n$.

The bilayer graphene films were prepared on top of an $n^{++}-$doped Si substrate covered with $\approx$ 300~nm layer of SiO$_2$ by the usual
mechanical exfoliation technique, and identified with Raman spectroscopy. The double-gated devices were prepared in the same way as outlined in
Ref[14]. A micrograph of a typical device is shown in Fig.~1a. Au leads were first defined by standard e-beam lithography technique. To form a
top-gate dielectric layer, a thin layer of polymethyl methacrylate (PMMA, MW 950K, 3\% Chlorobenzene) was spun on to the substrate at 6000 rpm
for 35 sec. PMMA on top of the flake was cross-linked by exposure of 30 KeV electrons at a dose of 21000~$\mu$C/cm$^2$. Finally, a 40 nm thick
Au gate was evaporated on top of the cross-linked PMMA covering the flake fully. A schematic of the vertical cross-section of the devices is
shown in Fig.~1a. The net carrier density on the BLG flakes is then given by $n = n_0 + \epsilon_{ox}V_{bg}/ed_{ox} +
\epsilon_{cp}V_{tg}/ed_{cp}$, where $n_0$ is the intrinsic doping, $\epsilon_{ox}$ and $d_{ox}$ are respectively the dielectric permittivity and
thickness of the SiO$_2$ layer, while $\epsilon_{cp}$ and $d_{cp}$ are those for the cross-linked polymer layer. The voltage applied on the back
(doped silicon) and top (gold) gates are denoted as $V_{bg}$ and $V_{tg}$, respectively. Typically, $d_{ox} \approx 300$~ nm and $d_{cp} \approx
100$~nm were used which made the topgate about three times more effective in inducing carriers in the BLG devices than the backgate
($\epsilon_{ox} \approx 4$, $\epsilon_{cp} \approx 4.5$). For the device presented in this paper, this was confirmed from the slope
$dV_{tg}/dV_{bg}\approx 0.32$ (Fig.~1d) which tracks the shift in the resistance maximum occurring at the overall charge neutrality when both
$V_{tg}$ and $V_{bg}$ are varied. The charge mobility of the device was estimated to be $\sim 1160$~cm$^2$/Vs, which contained an intrinsic hole
doping of $-n_0 \approx 5.82\times10^{11}$~cm$^{-2}$.

\begin{figure}
\begin{center}
\includegraphics[width=7.5cm,height=5cm]{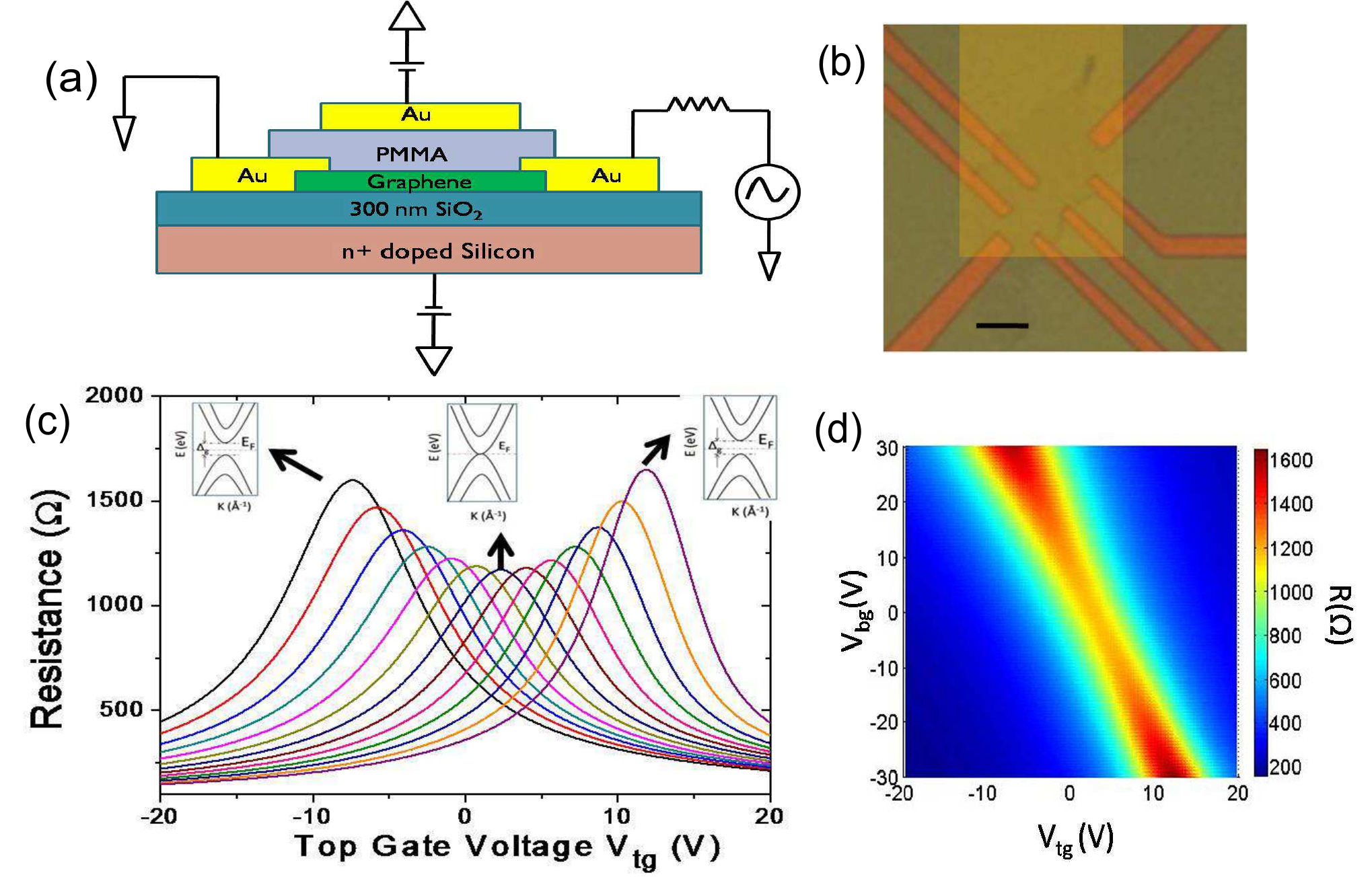}
\end{center}
\vspace{-0.7cm}\caption{ (a) Schematic of the double gated BLG device used in the experiment. (b) Optical micrograph of the device fabricated on
the flake. The scale bar is 10 $\mu$m. (c) Resistance versus topgate voltage for various back gate voltages, ranging from $30$~V to $-30$~V
(left to right) with an interval of $5$~V. Insets show schematics of corresponding band structures. (d) 2D color plot of the resistance as a
function of both top- and back- gate voltages at T = 107K, showing that the position of the charge neutrality peak shifts with both gate
voltages according to the capacitance ratio.} \label{figure1}
\end{figure}

When $V_{bg}$ and $V_{tg}$ are different, a finite ${\cal E}$ is established between the electrodes. The maximum voltage difference was
restricted to $|V_{bg} - V_{tg}|_{max} \lesssim 50$~V to avoid a dielectric breakdown. Consequently, our experiments were carried out at a
slightly lower temperature ($T \approx 107$~K) to obtain a clearly observable signature of band gap opening on charge transport. The
resistance($R$)$-V_{tg}$ characteristics of the device is shown in Fig.~1c for several different values of $V_{bg}$ spanning between $-30$ to
$+30$~V. Existence of the electric field-induced band gap becomes increasingly prominent at higher $V_{bg}$ with increasing $R$ at charge
neutrality (corresponding $V_{tg}$ henceforth denoted as $V^{Rmax}_{tg}$). Fig.~1c also shows the $R-V_{tg}$ characteristics to be rather
symmetric about $V^{Rmax}_{tg}$ in our devices which indicates high degree of electron-hole symmetry, and confirms the material/quality of the
contacts to be satisfactory for reliable noise measurements \cite{gordon2}.

\begin{figure}
\begin{center}
\includegraphics[width=6.5cm,height=11cm]{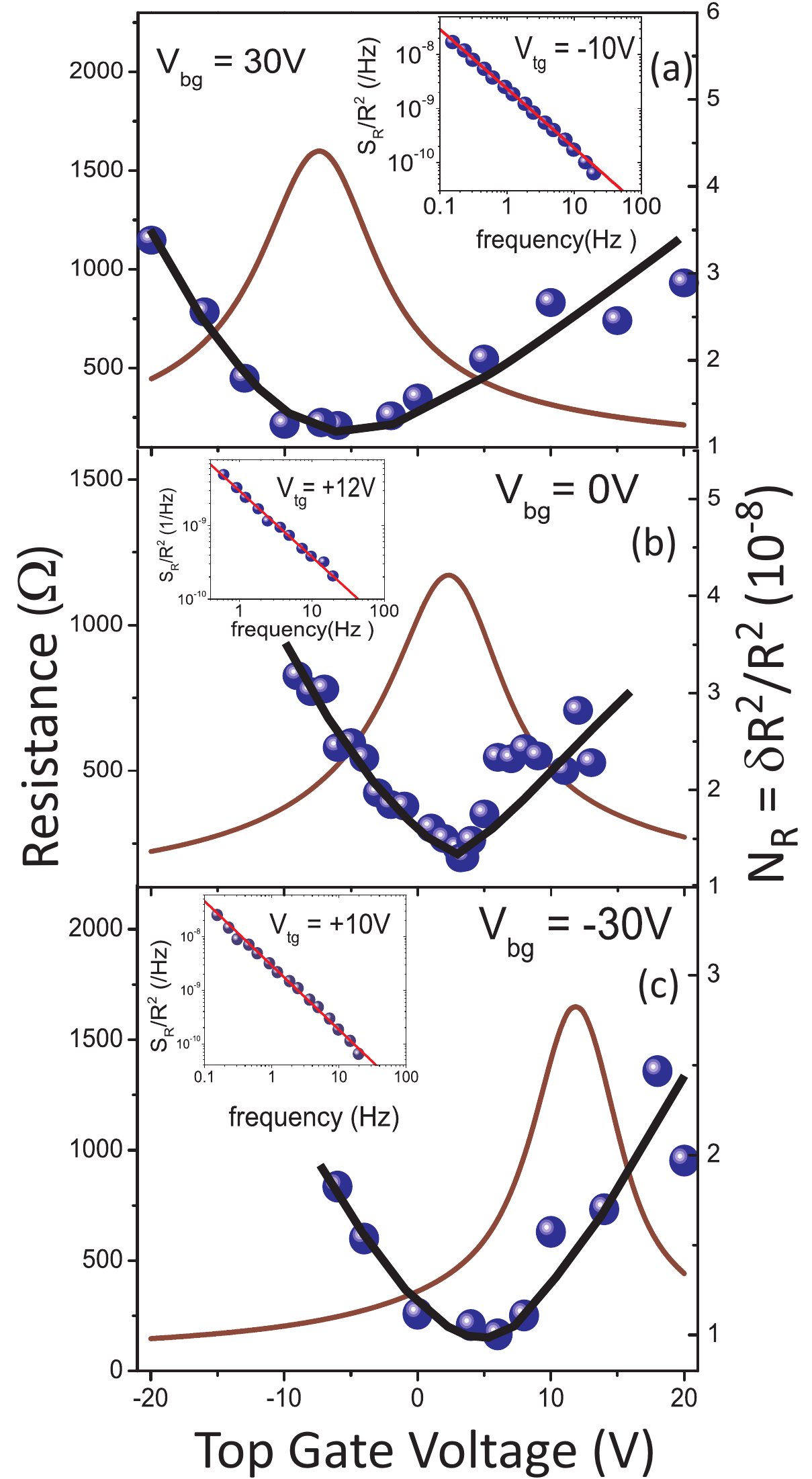}
\end{center}
\vspace{-0.7cm}\caption{Electrical transport and noise characterization of a
BLG device. The resistance and the normalized noise power spectral density
($N_R$) as functions of top gate voltages are shown for various back gate
voltages:(a) $30$~V (b) $0$~V (c) $-30$~V. The thick solid lines are guide to
the eye. The insets show typical noise power spectra $S_R/R^2$, far from the
charge neutrality point for each back gate voltage.} \label{figure2}
\end{figure}

Noise in the BLG devices were measured in low-frequency ac four-probe method, as well as in a five probe technique with a dynamically balanced
Wheatstone bridge, with both methods yielding similar results. Typical noise measurement involves digitization of the time-dependent output of
the lockin amplifier, followed by multistage decimation of the signal to eliminate effects of higher harmonic of the power line or other
unwanted frequencies, and finally estimation of the power spectral density $S_R(f)$ over a wide frequency bandwidth (See Ref[16] for details).
The excitation was below 500~nA to avoid heating and other non-linearities, and verified by quadratic excitation dependence of noise at a fixed
$R$. For all noise measurements, the voltages to both gates were provided from stacks of batteries instead of electronic voltage sources, which
resulted in a background noise level $< 1\times10^{-17}$~V$^2$/Hz, within a factor of $\sim 2$ of the Nyquist level. The background noise was
measured simultaneously, and subtracted from the total noise.

Typical power spectra of resistance noise are shown in the insets of Fig.~2a-c.
For comparison, the phenomenological Hooge relation provides a normalization
scheme for the resistance noise power spectral density as, $S_R(f) =
\gamma_HR^2/nA_Gf$, where the Hooge parameter $\gamma_H$ was found to be weakly
frequency dependent, and $A_G \approx 72\mu$m$^2$ is area of the BLG flake
between the voltage probes. Typically, we found $\gamma_H \sim 2\times10^{-3}$
sufficiently away from charge neutrality (at $n \sim
1.5\times10^{12}$~cm$^{-2}$), being similar to the value reported for BLG
nanoribbons~\cite{Avouris}. Here, however, instead of focusing on $\gamma_H$ or
noise magnitude at a specific frequency, we compute and analyze the total
variance of resistance fluctuations $N_R = \langle\delta R^2\rangle/R^2 =
(1/R^2)\int S(f)df$, which is essentially the normalized noise power spectral
density integrated over the experimental bandwidth.

Figs.~2a-c show the variation of $N_R$ and the corresponding average resistance
as functions of $V_{tg}$ at three different values of $V_{bg}$. For all
$V_{bg}$, $N_R$ shows a minimum at a specific $V_{tg}$, denoted as
$V^{Nmin}_{tg}$, and increases monotonically on both sides of $V^{Nmin}_{tg}$.
Similar behavior was observed for noise in BLG nanoribbons as
well~\cite{Avouris}, which confirms this to be an intrinsic characteristic of
BLG, although $V^{Rmax}_{tg}$ and $V^{Nmin}_{tg}$ are not equal in our
measurements, indicating that noise minimum has been shifted away from charge
neutrality in the presence of finite ${\cal E}$. In Fig.~3a, we have plotted
both $V^{Nmin}_{tg}$ obtained from noise measurements, as well as the
corresponding $V^{Rmax}_{tg}$, at various $V_{bg}$. As shown in Fig.~3b, this
allows us to follow the noise minimum point jointly with $n = n_\Delta =
(\epsilon_0\epsilon_{cp}/ed_{cp})(V^{Nmin}_{tg} - V^{Rmax}_{tg})$, and external
electric field ${\cal E} = {\cal E}_0 = (V_{bg} -
V_{tg}^{Nmin})/(d_{ox}+d_{cp})$.

\begin{figure}
\begin{center}
\includegraphics[width=8.5cm,height=4cm]{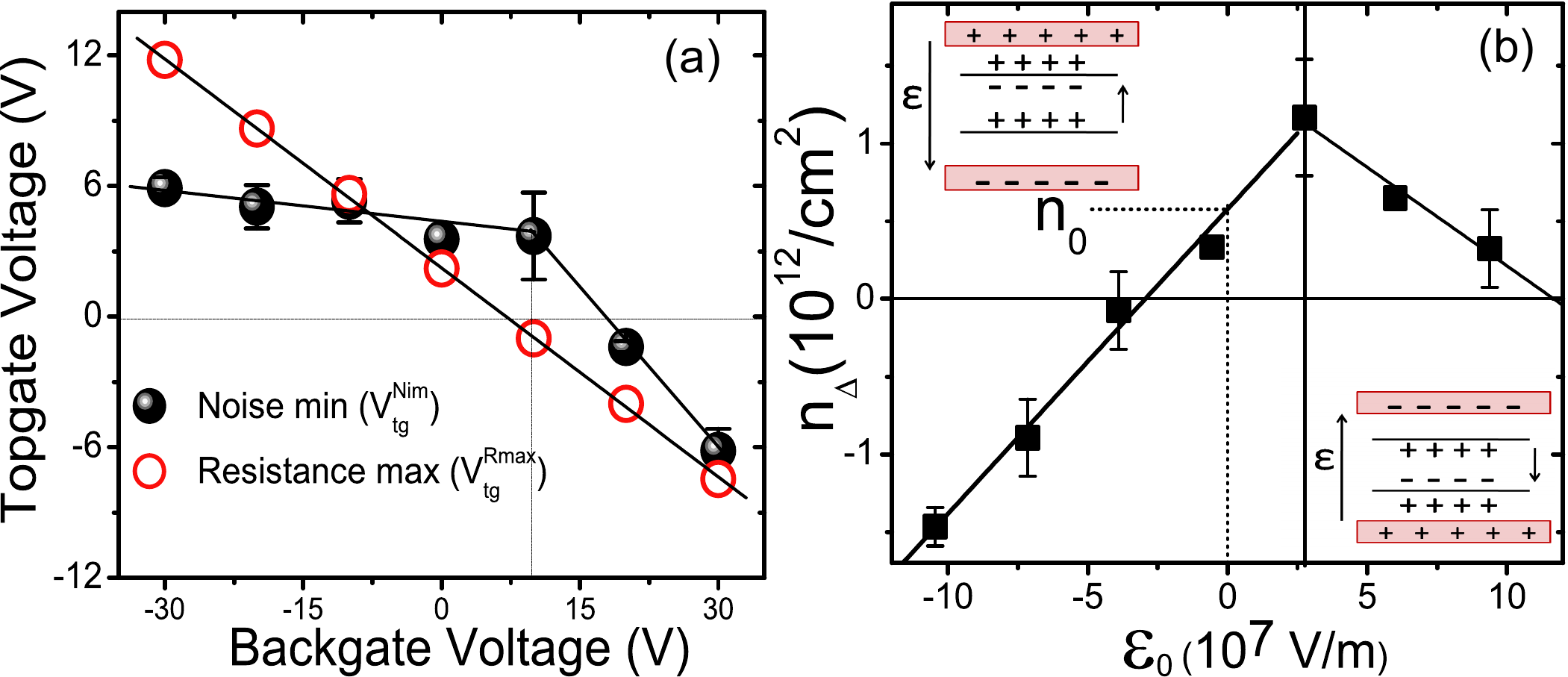}
\end{center}
\vspace{-0.7cm}\caption{(a) Top gate voltages at charge neutrality point
($V^{Rmax}_{tg}$) and noise minimum point($V^{Nmin}_{tg}$), plotted as
functions of the back gate voltages, extracted from Fig.~1 and Fig.~2. (b)
Dependence of charge density ($n_\Delta$) on external electric field (${\cal
E}_0$) at the noise minimum point. $n_\Delta$ was calculated from the
difference between the $V^{Rmax}_{tg}$ and $V^{Nmin}_{tg}$ (see text).}
\label{figure3}
\end{figure}

Two features in the dependence of $n_\Delta$ on ${\cal E}_0$ provide initial
indication that minimum of $N_R$ correspond to $\Delta_g = 0$. First, the
analysis with self-consistent Hartree interaction of Ref.~\cite{mccann2}, which
takes into account imperfect screening as well, shows that $n_\Delta$ and
${\cal E}_0$ are linearly related when $\Delta_g = 0$, as indeed observed
experimentally. Secondly, the vertical dashed line at ${\cal E}_0 = 0$ identify
that a minimum in $N_R$ occurs when the system is electron-doped by the same
amount as the intrinsic hole doping ($\approx n_0$), i.e., when the system is
charge neutral, which also corresponds to $\Delta_g = 0$.

To realize the implications of Fig.~3b quantitatively, we consider the case of
unscreened BLG, where the gates induce equal charge densities at the BLG layers
resulting in an interlayer electric field $= e(n_\Delta -
n_0)/2\epsilon_0\epsilon_r$ at $n = n_\Delta$, where $\epsilon_r$ is the
dielectric constant of the BLG region. The trap states in the substrate capture
charge from BLG which have two effects: First, intrinsic doping of the BLG by
$-n_0$ giving an additional inter-layer field $= -en_0/2\epsilon_0\epsilon_r$,
and secondly, the charged trap states would modify ${\cal E}_0$ by ${\cal E}_s
= en_0/\epsilon_0\epsilon_{ox}$. Hence, the variation of $n_\Delta$ with ${\cal
E}_0$ in Fig.~3b can be deduced by setting the total electric field between the
layers to be zero,

\begin{equation}
\label{eq1} |{\cal E}_0 - {\cal E}_s| + \frac{e(n_\Delta -
2n_0)}{2\epsilon_0\epsilon_r} = 0.
\end{equation}

\noindent Apart from confirming the bandgap to be zero at the minimum of $N_R$, Fig.~3b and Eq.~\ref{eq1} provide several crucial insights: (1)
{\it Charge organization:} Although we keep $V_{bg}$ fixed and vary only $V_{tg}$ to attain $n = n_\Delta$ in all cases, the charge organization
seems independent of this choice. Consequently, as shown in the insets of Fig.~3b, directionally opposite effective electric fields can be
screened at the same $n_\Delta$ by simply assuming the same doping with the opposite gate. This equivalence of the gates in spite of the
difference in relative orientation with respect to the BLG is in keeping with the suggestion that instead of a collection of two independent
capacitor plates, the BLG must be viewed as a single active component~\cite{Avouris}, which in our case is electrostatically coupled to an
assembly of two metallic electrodes. (2) {\it Screening:} In the presence of finite screening Eq.~\ref{eq1} continues to be valid since
screening modifies both external electric field and the excess charge density, leaving the linearity of $n_\Delta$ vs. ${\cal E}_0$ at $\Delta_g
= 0$ unaffected~\cite{mccann2}. (3) {\it Dielectric constant:} The slopes of the lines on either side of ${\cal E}_0 = {\cal E}_s$ allow direct
evaluation of the dielectric constant ($\epsilon_r$) which we find to be $\epsilon_r \approx 1.7\pm0.2$ and $\approx 1.2\pm0.2$ for two opposite
directions of effective external electric field (see insets of Fig.~3b). While these values are reasonable~\cite{castroneto}, the difference may
be due to the asymmetry in screening of disorder at the two configurations of charge distribution (insets of Fig.~3b).

\begin{figure}
\begin{center}
\includegraphics[width=8.5cm,height=7cm]{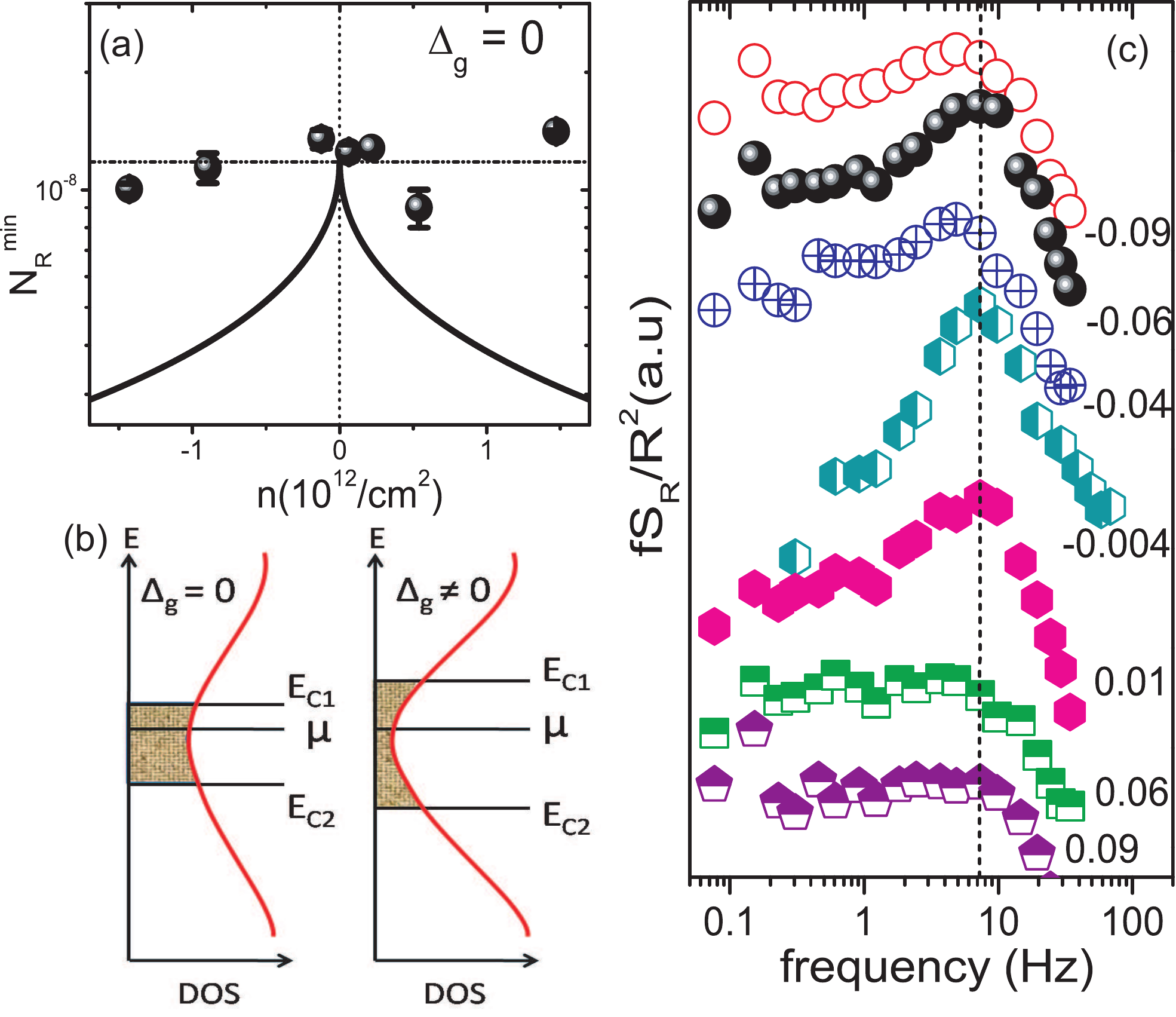}
\end{center}
\vspace{-0.7cm}\caption{(a) Variation of $N_{min}^R$ with $n$.The solid curve shows the calculated noise as a function of $n$ (see Ref[17]). (b)
Schematic representation shows the evolution of localization length $\xi$ with increasing gap, where $\xi \sim \hbar/(2m^*|E_C - \mu|)^{1/2}$.
$E_{C1}$ and $E_{C2}$ are the mobility edges, which increase with the band gap in our case, and $\mu$, the chemical potential, is insensitive to
the gate voltage due to proximity of the trap states. (c) $fS(f)$ is plotted as a function of frequency for various electric fields (V/nm),
which shows existence of a peak at $f \approx 7$~Hz.} \label{figure5}
\end{figure}

The important remaining question concerns the microscopic mechanism of
resistance noise in BLG, and in particular, the role of charge traps in the
substrate. In connection to this we have highlighted two aspects of BLG noise
in Fig.~4. At first, the magnitude of $N_R^{min}$ obtained at various $V_{bg}$
was found to be surprisingly insensitive to corresponding $n_\Delta$ (Fig.~4a).
Since $\Delta_g = 0$ at the minimum of noise, this insensitivity suggests that
the screening mechanism of external potential fluctuations remains largely
unaffected as the charge density is varied at least over the range of $|n|
\lesssim 10^{12}$~cm$^{-2}$, deviating significantly from simple calculations
based on screened Coulomb potential in BLG and delocalized
carriers~\cite{sdsarma}. To understand this we note that screening of an
external potential is determined by the $n$-dependence of local chemical
potential, but in the proximity of charge traps, discrete charging/discharging
of the trap states {\it locks} the chemical potential, making it insensitive to
the gate voltage. Such a phenomenon has been observed in compressibility
measurements in two dimensional systems close to the localization transition,
and attributed to dopant-related trap states~\cite{yacoby}. Hence the weak
variation in $N_R$ also suggests that quasiparticles in our BLG device are
nearly or weakly localized. Indeed, arbitrarily weak impurity potential has
been suggested to smear the bands and localize the quasiparticles at low
energies~\cite{nilsson}, particularly at the band tails, and experimental
evidence of localized states in electrically biased BLG has been observed in
low-temperature transport~\cite{oostinga}. In our devices as well, we found the
resistance at the charge neutrality point to increases with decreasing $T$
irrespective of ${\cal E}$ (not shown), suggesting quasiparticles to be in the
localized regime.

Among the models of low-frequency $1/f$-noise in systems with localized
carriers, where charge transport takes place through nearest-neighbor (high
$T$) or variable-range hopping (low $T$), both charge number fluctuations in
the percolating cluster~\cite{shklovskii} or energy-level modulation in the
hopping sites~\cite{kozub,savchenko2} have been discussed extensively. We
believe both mechanisms are effective in our BLG devices: (1) The sensitivity
of noise magnitude on screening indicates contribution from the energy-level
modulation mechanism, which involves sites that do not belong to the
percolation cluster. With increasing $\Delta_g$ the localization length $\xi$
decreases, which results in weaker screening and higher resistance noise (see
schematic of Fig.~4b.) (2) Secondly, theoretical framework of noise from a slow
charge exchange process between the percolation cluster and the trap states
predict a saturation of the power spectral density at very low
frequencies~\cite{shklovskii,shklovskii2}. To verify this we have plotted
$fS(f)$ as a function $f$ in Fig.~4c at $\Delta_g = 0$ for various ${\cal
E}_0$. Evidently, the low-frequency saturation in $S(f)$ manifests as peaks in
$fS(f)$ indicating the crossover frequency $\sim \nu_0\exp(-\sqrt{2/\pi
n_0\xi^2})$ to be $\approx 7$~Hz, where $\nu_0 \sim 10^{13}$~Hz. This gives a
reasonable estimate of $\xi \sim 0.5 - 0.6$~nm, within a factor of two of the
effective Bohr radius of BLG. It is unclear why the charge-exchange mechanism
is best detectable for low $|{\cal E}_0|$, but its experimental signature,
along with the locking of the chemical potential at low densities, provide a
consistent framework to understand the role of disorder in the electronic and
thermodynamic of BLG.

In conclusion, we have measured the low-frequency resistance noise in bilayer
graphene flakes as a function of charge density and inter-electrode electric
field. The absolute magnitude of noise is intimately connected with the BLG
band structure, and shows a minimum when the band gap of the system is zero.
The experiments also reveal the charge organization in BLG-based electronic
devices, and the microscopic mechanism of resistance noise.

We acknowledge the support from Institute Nanoscience Initiative, Indian Institute of Science, Bangalore 560 012, India.


\begin{references}
\bibitem{mccann1} E. McCann and V. I. Fal'ko, Phys. Rev. Lett. \textbf{96}, 086805 (2006).
\bibitem{mccann2} E. McCann, Phys. Rev. B \textbf{74}, 161403(R) (2006).
\bibitem{castroneto} E. V. Castro \emph{et al}., e-print arXiv:condmat/0807.3348v1 (2008).
\bibitem{castro} E. V. Castro \emph{et al}., Phys. Rev. Lett. \textbf{99}, 216802 (2007).
\bibitem{ohta} T. Ohta \emph{et al}., Science \textbf{313}, 951 (2006).
\bibitem{oostinga} J. B. Oostinga \emph{et al}., Nature Mater. \textbf{7}, 151 (2008).
\bibitem{stauber} T. Stauber, N. M. R. Peres, F. Guinea, and A. H. Castro Neto, Phys. Rev. B \textbf{75}, 115425 (2007).
\bibitem{savchenko} R.V. Gorbachev \emph{et al}., Phys. Rev. Lett. \textbf{98}, 176805 (2007).
\bibitem{koshino} M. Koshino, Phys. Rev. B \textbf{78}, 155411 (2008).
\bibitem{tailstate} V. V. Mkhitaryan and M. E. Raikh, e-print arXiv:condmat/0807.2445v1 (2008).
\bibitem{Avouris} Y. Lin and Phaedon Avouris, Nano Lett. \textbf{8}, 2119 (2008).
\bibitem{Avouris2} Y. Lin, J. Appenzeller, J. Knoch, Z. Chen, and P. Avouris, Nano Lett. \textbf{6}, 930 (2006).
\bibitem{sahu} H. Min, B. Sahu, S. K. Banerjee, and A. H. MacDonald, Phys. Rev. B \textbf{75}, 155115 (2007).
\bibitem{gordon} D. Goldhaber-Gordon \emph{et al}., Phys. Rev. Lett. \textbf{98}, 236803 (2007).
\bibitem{gordon2} B. Huard, N. Stander, J. A. Sulpizio, and D. Goldhaber-Gordon, Phys. Rev. B \textbf{78}, 121402(R) (2008).
\bibitem{arindam} A. Ghosh and A. K. Raychaudhuri, Phys. Rev. Lett. \textbf{84}, 4681 (2000); A. Ghosh \emph{et al}., e-print arXiv:condmat/0402130 v1 (2004).
\bibitem{sdsarma} E. H. Hwang and S. Das Sarma, e-print arXiv:condmat/0804.2255v1 (2008). Noise is estimated in {\it local interference} model with a
screened Coulomb potential $\phi(r) \sim \sin(2k_Fr)/(2k_Fr + Cq_{TF}r)^2$, at large distances $r$, where $q_{TF}$ is Thomas-Fermi screening
wave vector, and $C$ is a numerical constant.
\bibitem{yacoby} S. Ilani, A. Yacoby, D. Mahalu, and Hadas Shtrikman, Phys. Rev. Lett. \textbf{84}, 3133 (2000).
\bibitem{nilsson} J. Nilsson and A. H. Castro Neto, Phys. Rev. Lett. \textbf{98}, 126801 (2007).
\bibitem{shklovskii} B.I. Shklovskii, Solid State Commun. \textbf{33}, 273 (1980).
\bibitem{kozub} V.I. Kozub, Solid State Commun.\textbf{97}, 843 (1996).
\bibitem{savchenko2} V. Ya. Pokrovskii, A. K. Savchenko, W. R. Tribe, and E. H. Linfield, Phys. Rev. B \textbf{64}, 201318(R) (2001).
\bibitem{shklovskii2} B.I. Shklovskii, Phys. Rev. B \textbf{67}, 045201 (2001).


\end{references}
\end{document}